\documentstyle[amssymb]{article}
\font\tenrm=cmr10 \font\tenit=cmti10 \font\elevenbf=cmbx10
scaled\magstep 1 \font\elevenrm=cmr10 scaled\magstep 1
\newcommand{\be}{\begin{equation}}
\newcommand{\ee}{\end{equation}}

\newcommand{\bea}{\begin{eqnarray}}
\newcommand{\eea}{\end{eqnarray}}

\newcommand{\al}{\alpha}

\newcommand{\pr}{\partial}

\newcommand{\dg}{\dagger}

\input{epsf}
\renewenvironment{thebibliography}[1]
 { \elevenrm
   \begin{list}{\arabic{enumi}.}
    {\usecounter{enumi}     \setlength{\parsep}{0pt}
     \setlength{\itemsep}{3pt} \settowidth{\labelwidth}{#1.}
     \sloppy
    }}
     {\end{list}}
\begin{document}

\begin{center}
{\elevenbf On the Q-Ball Profile Function}\\[0pt]
\vglue0.5cm {\tenrm T.A. Ioannidou\footnote{{\it Permanent
Address}:
Institute of Mathematics, University of Kent, Canterbury CT2 7NF, UK}$%
^{\dagger }$ and N.D. Vlachos$^{\ddagger }$ } \\[0pt]
\vglue0.3cm $^{\dg}${\tenit Mathematics Division, School of Technology,
University of Thessaloniki, Thessaloniki 54124, Greece}\\[0pt]
$^{\ddagger }${\tenit Physics Department, University of Thessaloniki,
Thessaloniki 54124, Greece }\\[0pt]
\vglue0.3cm {\it Emails: T.Ioannidou@ukc.ac.uk\\[0pt]
\hspace{12mm} vlachos@physics.auth.gr}
\end{center}

\vglue 0.3cm {\rightskip=2pc  \leftskip=2pc
\tenrm\baselineskip=11pt \noindent We use analytic and numerical
methods  to obtain  the solution of the Q-ball equation of motion.
In particular, we show that the profile function of the
three-dimensional Q-ball can be accurately approximated by the
symmetrized Woods-Saxon distribution. \vglue 0.6cm} \elevenrm \baselineskip%
=14pt

\section{Introduction}

The existence of Q-balls is a general feature of scalar field
theories carrying a conserved $U(1)$ charge \cite{1,Lee}. Q-balls
can be understood as bound states of scalar particles and appear
as stable classical solutions (nontopological solitons) carrying a
rotating time dependent internal phase. They are characterized by
a conserved nontopological charge $Q$ (Noether charge) which is
responsible for their stability (see, for example, Refs.
\cite{drohm}-\cite{belova}). These features differentiate the
Q-ball interaction properties from those of the topological
solitons since here the charge $Q$ can take arbitrary values in a
specific range, allowing for the possibility of charge transfer
between solitons during the interaction process.

Up till now, comprehensive studies of these objects have been made
by using either numerical simulations \cite{drohm}-\cite{AKP} or
some analytic considerations \cite{1,MV,PC}. Recently, in
\cite{IKV} the explicit relation between the energy and the charge
of the Q-balls has been derived using analytic arguments. In
particular, a  semi-Bogomolny argument in the energy density led
to a first order differential equation whose solution however, did
not satisfy the correct boundary conditions and differed
considerably from its exact form. In this work, we present a
method to obtain an analytic form  for the Q-ball profile function
which is in good agreement with the numerical results.

We consider the $U(1)$ Goldstone model, given by the Lagrangian
\begin{equation}
{\cal L}=\frac{1}{2}\partial _{\mu }\phi \,\partial ^{\mu }\bar{\phi}%
-U(|\phi |)  \label{L}
\end{equation}%
where $\phi $ is a single complex scalar field in three spatial dimensions
while the potential $U(|\phi |)$ is a function  of $|\phi |$ only and has a
single minimum at $\phi =0$. This is equivalent of stating that there is a
sector of scalar particles (mesons) which carry $U(1)$ charge and have mass
squared equal to $\frac{1}{2}U^{\prime \prime }(0)$. The corresponding
energy functional is given by
\begin{equation}
E=\int \left( \frac{1}{2}|\pr_t{\phi}|^{2}+\frac{1}{2}|\nabla \phi
|^{2}+U(|\phi |)\right) d^{3}x.  \label{En}
\end{equation}%
The model has a global $U(1)$ symmetry and an associated conserved Noether
current $J_{\mu }$ exists whose covariant conservation $\partial ^{\mu
}J_{\mu }=0$ leads to the existence of the conserved Noether charge $Q$
given by
\begin{equation}
Q=\frac{1}{2i}\int \left( \bar{\phi}\,\partial _{t}\phi -\phi \,\partial _{t}%
\bar{\phi}\right) d^{3}x.
\end{equation}%
A stationary Q-ball solution has the form
\begin{equation}
\phi =e^{i\omega t}f(r)  \label{ph}
\end{equation}%
where $f(r)$ is a real radial profile function which satisfies the ordinary
differential equation
\begin{equation}
f^{\prime \prime }(r)+\frac{2}{r}f^{\prime }(r)=-\omega ^{2}f(r)+U^{\prime
}(f)  \label{gene}
\end{equation}%
with the condition $f(\infty )=0$ and $f^{\prime }(0)=0$. This equation can
either be interpreted as describing the motion of a point particle moving in
a potential with friction \cite{1}, or in terms of Euclidean bounce
solutions \cite{23}. In each case the effective potential being $%
U_{eff}(f)=\omega ^{2}f^{2}/2-U(f)$ leads to constraints on the potential $%
U(f)$ and the frequency $\omega $ in order for a Q-ball solution to exist.
Firstly, the effective mass of $f$ must be negative. If we consider a
potential $U(f)$ which is non-negative and satisfies $U(0)=U^{\prime }(0)=0$%
, $U^{\prime \prime }(0)=\omega _{+}^{2}>0$ then one can deduce that $\omega
<\omega _{+}$. Furthermore, the minimum of $U(f)/f^{2}$ must be attained at
some positive value of $f$, say $0<f_{0}<\infty $ and existence of the
solution requires that $\omega >\omega _{-}$ where $\omega
_{-}^{2}=2U(f_{0})/f_{0}^{2}$. Hence, Q-balls exist for all $\omega $ in the
range $\omega _{-}<|\omega |<\omega _{+}$.

Then, the charge and the energy of a stationary Q-ball solution simplify to
\begin{eqnarray}
Q &=&4\pi \omega \int r^{2}\,f^{2}(r)\,dr  \label{Q} \\
E &=&4\pi \int \left( \frac{1}{2}\,\omega ^{2}f^{2}(r)+\frac{1}{2}%
\,f^{\prime }{}^{2}(r)+U(f)\right) r^{2}\,dr.  \label{E}
\end{eqnarray}%
It has been observed using numerical and analytic methods that the classical
stability of a Q-ball is related with the dependence of its charge on the
internal frequency $\omega $. For small internal frequency, close to its
minimal value $\omega _{-}$, the profile function is almost constant which
implies that the charge (\ref{Q}) is large and this corresponds to the
so-called thin-wall approximation. On the other hand, for large internal
frequency (close to its maximal value $\omega _{+}$) the profile function
(and thus the charge) tends to zero and this corresponds to the thick-wall
approximation. In fact, for $\omega \rightarrow \omega _{+}$, the behavior
of the charge $Q$ depends on the particular form of the potential and the
number of dimensions \cite{PC}. In the case studied here we show that $%
Q\rightarrow \infty $ as $\omega \rightarrow \omega _{+}$.

The choice of the potential is not unique; the standard
requirement is that the function $U(f)/f^2$ has a local minimum at
some value of $f$ different from zero. For simplicity, we consider
the following form
\begin{equation}
U(f)=f^2\left(1+(1-f^2)^2\right)  \label{Uf}
\end{equation}
which implies that in terms of the earlier notation we have that $\omega_+=2$
and $\omega_-=\sqrt{2}$, and therefore  stable Q-balls exist for $\sqrt{2}%
<\omega<2$.  In this case, the energy-charge dependence of the
Q-balls, obtained in \cite{IKV} using a semi-Bogomolny argument,
is given by the analytic expression
\begin{eqnarray}
E_{_{Bog}}\!\!\!\!&=&\!\!\!\!\sqrt{2}\,Q_{_{Bog}}+\frac{3^{2/3} \pi^{1/3}}{%
2^{7/6}}\, Q_{_{Bog}}^{2/3}+ \frac{5 \pi^{2/3}}{2^{11/6}3^{2/3}}%
\,Q_{_{Bog}}^{1/3}-\frac{\pi \left( 4+3\pi^{2}\right) }{36\sqrt{2}}+ \frac{%
\pi ^{4/3}\left( 17-216 \pi^{2}\right) }{2592~2^{1/6}~3^{1/3}}\
Q_{_{Bog}}^{-1/3}  \nonumber \\
\!\!\!\!&&\!\!\!\!+\frac{\pi ^{5/3}\left( 20-54\pi ^{2} +27\pi^{4}\right) }{
1944~2^{5/6}~3^{2/3}}\,Q_{_{Bog}}^{-2/3} +O\left(Q_{_{Bog}}^{-1}\right).
\label{EQ}
\end{eqnarray}
For large $Q$, the above equation corresponds to the upper energy bound.
The same result can be obtained by representing the Q-ball profile function
by the Woods-Saxon distribution  (ie a generalization of the semi-Bogomolny
solution) as presented in \cite{IKV}.

Let us emphasize that neither the semi-Bogomolny nor the
Woods-Saxon {\it ansatz} describe accurately the Q-ball profile
function.  In particular, in both cases the derivative of the
profile function at the origin is non-zero (that is, different
from the required boundary conditions) while the corresponding
profile function differs from the exact one obtained by solving
numerically the equation of motion.  In the next section, we show
that the symmetrized Woods-Saxon distribution describes accurately
the Q-ball profile function and satisfies the correct boundary
conditions.

\section{Equation of Motion}

The Q-ball equation of motion (\ref{gene}) for the specific potential (\ref%
{Uf}) is given by
\begin{equation}
f^{\prime \prime }(r)+\frac{2}{r}\,f^{\prime }(r)=\left( 4-\omega
^{2}\right) f(r)-8f^{3}(r)+6f^{5}(r)  \label{prof}
\end{equation}%
and satisfies the boundary conditions $f^{\prime }(0)=0$,
$f(\infty )=0$. For large values of the argument $r$  the
nonlinear terms in (\ref{prof})
can be neglected and we get the asymptotic behaviour $f(r)\sim \exp (-\sqrt{%
4-\omega ^{2}}\,r).$ The exact solution of (\ref{prof}) should contain only
one free parameter $\omega $ so that the charge $Q$ the energy $E$ and the
initial value $f(0)$ should in principle be expressed as functions of this
parameter. Since equation (\ref{prof}) is too complicated to be tackled by
analytic methods, we shall seek a suitable {\it ansatz }for the profile
function and search for  relations connecting $E$,$Q$ and $f(0)$ with $%
\omega $.

As mentioned earlier, a test profile of the  Woods-Saxon type
cannot reasonably approximate the Q-ball solution since it fails
to satisfy the correct boundary conditions. Nevertheless, it
implies an energy-charge relation (\ref{EQ}) which holds
remarkably well for a wide range of energies, a fact that
certainly requires  some further investigation. A more general
test function that satisfies the correct boundary conditions and
has the correct asymptotic behaviour is given by the symmetrized
Woods-Saxon distribution
\begin{equation}
f(r)=\frac{c}{\sqrt{1+c_{1}\cosh \left( \alpha r\right) }}\ \cdot
\label{WSsym}
\end{equation}%
The values of the arbitrary parameters  $c$, $c_{1}$ and $\alpha $ can then
be  determined by fitting the data of the numerically solved (\ref{prof}%
). Having done that, we got a very satisfactory agreement in all
cases.  It is interesting to realize that $f(r)$  satisfies the
following differential equation
\begin{equation}
f^{\prime \prime }(r)+\frac{2}{r}f^{\prime }(r)=\frac{\alpha ^{2}}{4}\left(
1-\frac{4}{ar}\right) f(r)-\frac{\alpha ^{2}}{c^{2}}\left( 1-\frac{1}{ar}%
\right) f^{3}(r)+\frac{3\alpha ^{2}}{4c^{4}}\left( (1-c_{1}^{2})+\frac{2}{3}%
\frac{c_{1}^{2}}{\alpha r}~\right) f^{5}(r)+O\left( f^{7}\right)
\label{sym}
\end{equation}%
which in the limit  $ar\gg 4$ looks exactly like (\ref{prof}). It
is then quite reasonable to expect that there must be a critical
value of $r=r_{c}$ beyond which the following equations must
approximately hold true:
\begin{eqnarray}
&&\frac{\alpha ^{2}}{4}\left( 1-\frac{4}{\alpha r_{c}}\right) \simeq
4-\omega ^{2} \\
&&\frac{\alpha ^{2}}{c^{2}}\left( 1-\frac{1}{\alpha r_{c}}\right) \simeq 8 \\
&&\frac{3\alpha ^{2}}{4c^{4}}\left( (1-c_{1}^{2})+\frac{2}{3}\frac{c_{1}^{2}%
}{\alpha r_{c}}~\right) \simeq 6\ \cdot   \label{16}
\end{eqnarray}%
The thin-wall approximation is reached when $c_{1}\rightarrow 0$
so we will assume that $c_{1}<1$ and we shall neglect  the last
term in the left hand side of (\ref{16}). Then the system above
gives the approximate solutions
\begin{eqnarray}
c &\simeq &\sqrt{\frac{2}{3}}\sqrt{1-c_{1}^{2}}\sqrt{1+\sqrt{1-\frac{3}{8}%
\frac{4-\omega ^{2}}{1-c_{1}^{2}}}}  \label{c} \\[3mm]
\alpha  &\simeq &\frac{4\sqrt{2}}{3}\sqrt{1-c_{1}^{2}}\left( 1+\sqrt{1-\frac{%
3}{8}\frac{4-\omega ^{2}}{1-c_{1}^{2}}}\right)   \label{al} \\[3mm]
\alpha r_{_{c}} &\simeq &\frac{1+\sqrt{1-\frac{3}{8}\frac{4-\omega ^{2}}{%
1-c_{1}^{2}}}}{-\frac{1}{2}+\sqrt{1-\frac{3}{8}\frac{4-\omega ^{2}}{%
1-c_{1}^{2}}}}\ \cdot   \label{ar}
\end{eqnarray}
A comparison with values obtained by a direct fit shows that the
relations above are quite good. In fact, equation (\ref{c}) is
excellent, while equation (\ref{al}) gives the right shape but the
actual values for $\al$ are on the average $10\%$ higher than
expected, implying a faster drop of the  profile function. Note
that both $c$ and $a$ are slowly varying functions of $\omega $.

The equation of motion (\ref{prof})  can be written as:
\begin{equation}
\frac{d}{dr}\left( \frac{1}{2}\,f^{\prime }{}^{2}(r)+\frac{1}{2}\,\omega
^{2}f^{2}(r)-U(f)\right) =-\frac{2}{r}\,f^{\prime }{}^{2}(r)\ \cdot
\label{od}
\end{equation}%
In the absence of the friction term, equation (\ref{od}) would
simply imply the conservation of energy for the corresponding
mechanical problem. In the presence of friction and upon
integrating (\ref{od})  we get that the initial potential energy
equals  the work done by friction. This relation can be used to
provide a further constraint on the form of (\ref{WSsym})
\begin{equation}
\frac{1}{2}\,\omega ^{2}f^{2}(0)-U\left( f(0)\right) =-2\int_{0}^{\infty }%
\frac{\,f^{\prime }{}^{2}(r)}{r}\,dr\ \cdot   \label{fric}
\end{equation}%
Recall that  $f(0)=c/\sqrt{1+c_1}$. Finally, using the symmetrized
Woods-Saxon distribution (\ref{WSsym}), the charge (\ref{Q}) and
the energy (\ref{E}) of the Q-ball can be explicitly evaluated in
terms of the parameters $c$, $c_{1}$ and $\alpha $:
\begin{eqnarray}
Q &=&4\pi \omega \,\frac{c^{2}}{3\alpha ^{3}\sqrt{1-c_{1}^{2}}}\,\cosh
^{-1}\left( \frac{1}{c_{1}}\right) \left( \pi ^{2}+\cosh ^{-1}\left( \frac{1%
}{c_{1}}\right) ^{2}\right) ,  \label{QWS} \\[3mm]
E &=&4\pi \left[ \frac{\alpha ^{2}+4\left( 4+\omega ^{2}\right) }{8}%
\,I_{0}-\left( 2+\frac{\alpha ^{2}}{4c^{2}}\right) I_{1}+\left( 1+\frac{%
\alpha ^{2}}{8c^{4}}\left( 1-c_{1}^{2}\right) \right) I_{2}\right] ,
\label{EWS}
\end{eqnarray}%
where the integrals $I_{0}$, $I_{1}$ and $I_{2}$ are given by
\begin{eqnarray}
I_{0} &=&\frac{Q}{4\pi \omega }, \\[3mm]
I_{1} &=&c^{2}I_{0}+c^{2}c_{1}\frac{d}{dc_{1}}I_{0}, \\[3mm]
I_{2} &=&c^{4}I_{0}+2c^{4}c_{1}\frac{d}{dc_{1}}I_{0}+\frac{1}{2}%
c^{4}c_{1}^{2}\frac{d^{2}}{dc_{1}^{2}}I_{0}\ \cdot
\end{eqnarray}%
It is readily seen that the actual values for $Q$ and $E$ depend strongly on
the value  of $c_{1}$ especially for small values of $c_{1}$. In this
region, $c_{1}$ can be eliminated in favour of $Q$ to get
\begin{eqnarray}
E &=&\left( \frac{1}{\omega }+\frac{\omega }{2}\right) Q+\frac{24+3\,\al^{2}}{%
16\,\omega }\left( q\omega \right) ^{\frac{1}{3}}\,{Q}^{\frac{2}{3}}+\frac{%
24+3\, \al^{2} }{8\omega
\,}\left(
q\omega \right) ^{\frac{2}{3}}\,{Q}^{\frac{1}{3}}
-\frac{\left( 8+\,\al^{2}\right) \,{\pi }^{2}}{8\,\omega }q\omega ~+{{O}(Q}
^{-\frac{1}{3}}{)\quad }  
\label{EQs}
\end{eqnarray}%
where%
\begin{equation}
q=\frac{4\pi c^{2}}{3\al^{3}}\ \cdot
\end{equation}%
It is interesting to see that the leading terms of  (\ref{EQs}) are
identical to those of (\ref{EQ}) for $\al=2\sqrt{2}$, $c=1$, $\omega =\sqrt{2}$%
. Taking into account the fact that $\al$ and $c$ are slowly
varying functions
of $\omega $ helps to explain the wide range of applicability of (\ref%
{EQ}).

As a result, the parameters $c$, $c_{1}$, $\alpha $ and $\omega $ for
specific values of the charge $Q$ can be uniquely determined by solving the
system of  equations: (\ref{c}), (\ref{fric}), (\ref{QWS}), and (\ref{EWS})
provided that the energy-charge dependence of the Q-ball is given by (\ref%
{EQ}). We expect the values obtained that way to be quite accurate
as long
as (\ref{EQ}) remains reasonably accurate. In practice, this means that $%
Q\gtrsim 16$, a value far distant from the values normally
associated with the region of the thin-wall approximation. In
Table 1 the values of the arbitrary constants in (\ref{WSsym}) are
presented for different values of $Q $ (or $\omega $). That way,
the initial value $f(0)$ can be determined explicitly; a
comparison with  values obtained numerically show that the
symmetrized Woods-Saxon distribution describes accurately the
Q-ball profile function.

\begin{center}
\begin{tabular}{|c|c|c|c|c|c|c|}
\hline $Q$ & $\omega $ & $c$ & $c_{1}$ & $\al$ & $f(0)$ & $f(0)_{num}$ \\
\hline 23.68 & 1.79 & 1.097 & 0.12 & 2.66 & 1.037 & 1.024
\\ \hline 35 & 1.72 & 1.087 & 0.068 & 2.60 & 1.052 & 1.055 \\
\hline 61.6 & 1.64 & 1.068 & 0.0243 & 2.54 & 1.056 & 1.065 \\
\hline 149.3 & 1.57 & 1.049 & 0.0031 & 2.53 & 1.048 & 1.056 \\
\hline
\end{tabular}
\newline
\end{center}

Table 1: Values of the parameters $c$, $c_1$, $\al$ and $f(0)$
obtained from $(\ref{WSsym})$ and numerically ($f(0)_{num}$) for
different values of  $\omega$.
\newline

\begin{figure}[tbp]
\begin{center}
\epsfxsize=9cm\epsfysize=9cm\epsffile{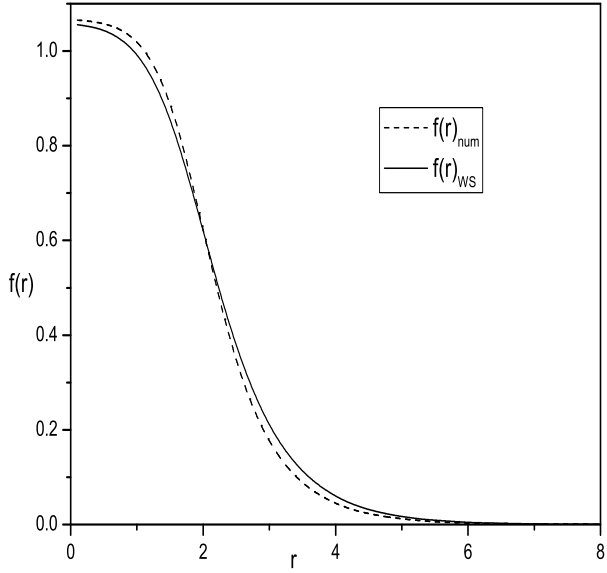}
\end{center}
\par
\vskip -1cm
\caption{The profile function $f(r)$ as a function of $r$ for $\protect\omega%
=1.64$} \label{fig-f(r)}
\end{figure}

\begin{figure}[tbp]
\begin{center}
\epsfxsize=9cm\epsfysize=9cm\epsffile{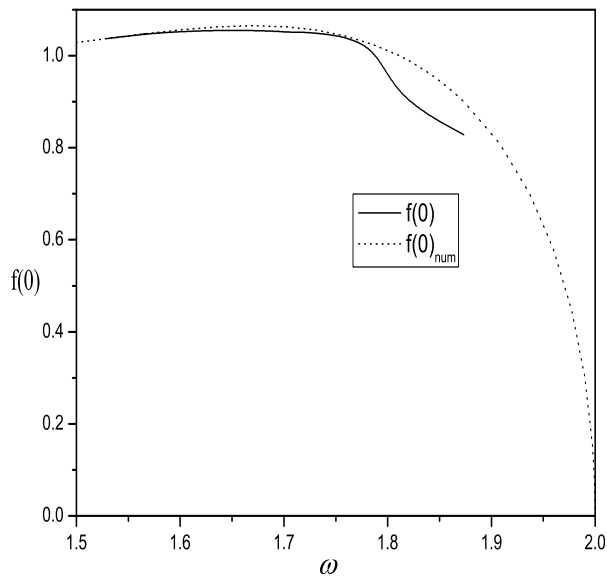}
\end{center}
\par
\vskip -1cm
\caption{The frequency dependence of the profile function at the origin $f(0)
$.}
\label{fig-f(0)}
\end{figure}

 Finally, Figure \ref{fig-f(r)} presents the profile function obtained
analytically and numerically for $\omega =1.64$. Figure
\ref{fig-f(0)}
presents the values of the profile function at the origin for different $%
\omega $ obtained numerically and analytically.

\section{Conclusions}

In this work, the basic properties of the Q-ball profile function
have been extensively studied by means of  mainly analytic
methods. In particular, it has been shown that the profile
function can be accurately approximated by the symmetrized
Woods-Saxon distribution, while the corresponding energy and
charge can be explicitly evaluated. The approach presented here
might prove to be particularly useful in understanding the basic
properties of Q-balls such as existence, small vibrations and
stability. We believe that a similar line of argument can be
applied to study  the profile function and the energy-charge
dependence in other types of potentials. In fact, we expect that
the symmetrized Woods-Saxon distribution will accurately describe
all types of Q-balls independently of the exact form of the scalar
potential. Work in this direction is currently in progress.

\section*{Acknowledgement}

We thank V. Kopeliovich, A. Kouiroukidis and S. Massen for useful
discussions. TI thanks the Royal Society and the National Hellenic Research
Foundation for a Study Visit grant.\newline

{\elevenbf\noindent References} \vglue 0.1cm


\begin{thebibliography}{99}
\bibitem{1}  S. Coleman, Nucl. Phys. B 262, 263 (1985)

\bibitem{Lee} T.D. Lee, Particle Physics and Introduction to Field Theory,
Harwood, London (1981)



\bibitem{drohm} J.K. Drohm, L.P. Yok, Y.A. Simonov, J.A. Tyon and V.I.
Veselov, Phys. Lett. B 101, 204 (1981)

\bibitem{belova} T.I. Belova and A.E. Kudryavtsev, JETP 68, 7 (1989)


\bibitem{AKP} M. Axenides, S. Komineas, L. Perivolaropoulos and M. Floratos,
Phys. Rev. D 61, 085006 (2000)

\bibitem{MV} T. Multamaki and I. Vilja, Nucl. Phys. B 574, 139 (2002);
hep-ph/9908446

\bibitem{PC} F. Paccetti Correia and M.G. Schmidt, Eur. Phys. J.
C 21, 181 (2001); hep-th/0103189

\bibitem{IKV} T. Ioannidou, V. Kopeliovich and N.D. Vlachos, {\it %
Energy-Charge Dependence for the Q-balls}, (2003); hep-th/0302013

\bibitem{23} A. Kusenko, Phys. Lett. B 404, 285 (1997)

\bibitem{Bk} H.T. Davis, Introduction to Nonlinear Differential and Integral
Equations, Dover Publications (1962)
\end{thebibliography}
\end{document}